\documentclass[preprint,showpacs,preprintnumbers,amsmath,amssymb]{revtex4}

\usepackage{graphicx}
\usepackage{dcolumn}
\usepackage{bm}
\usepackage{hyperref}

\begin{document}

\preprint{The following article has been submitted to Journal of Applied Physics (29th ICPS - invited paper)\newline}
\preprint{After it is published, it will be found at http://jap.aip.org/}

\title{Polarization fine-structure and enhanced single-photon emission of self-assembled lateral InGaAs quantum dot molecules embedded in a planar micro-cavity}

\author{C.~ Hermannst\"adter}
 \email{c.hermannstaedter@ihfg.uni-stuttgart.de}

\author{M.~Witzany}
\author{G.~J.~Beirne}
\author{W.-M.~Schulz}
\author{M.~Eichfelder}
\author{R.~Rossbach}
\author{M.~Jetter}
\author{P.~Michler}
\affiliation{Institut f\"ur Halbleiteroptik und Funktionelle Grenzfl\"achen, Universit\"at Stuttgart, Allmandring 3, 70569 Stuttgart, Germany}

\author{L.~Wang}
\affiliation{Max-Planck-Institut f\"ur Festk\"orperforschung, Heisenbergstr. 1, 70569 Stuttgart, Germany}

\author{A.~Rastelli}
\author{O.~G.~Schmidt}
\affiliation{Institut f\"ur Integrative Nanowissenschaften, Leibniz-Institut f\"ur Festk\"orper- und Werkstoffforschung, Postfach 27 01 16, 01171 Dresden, Germany}

\date{\today}

\begin{abstract}
Single lateral InGaAs quantum dot molecules have been embedded in a planar micro-cavity in order to increase the luminescence extraction efficiency. Using a combination of metal-organic vapor phase and molecular beam epitaxy samples could be produced that exhibit a 30 times enhanced single-photon emission rate. We also show that the single-photon emission is fully switchable between two different molecular excitonic recombination energies by applying a lateral electric field. Furthermore, the presence of a polarization fine-structure splitting of the molecular neutral excitonic states is reported which leads to two polarization-split classically correlated biexciton-exciton cascades. The fine-structure splitting is found to be on the order of 10 $\mu$eV.
\end{abstract}

\maketitle

\section{\label{sec:Introduction}Introduction}

Coupled quantum dot (QD) systems have been extensively studied during the past years primarily because they have the potential to combine atom-like single-dot properties with their specific coupling properties. Great progress has already been reported on the basis of QDs that are electrostatically defined in a two-dimensional electron gas. The charge state and electron spins have been coherently manipulated in double-dots \cite{Hayas,Petta,Koppe}, spin- and charge-readout was performed using the coupling of a QD to a quantum point contact \cite{Elzer} and the first quantum gates have been realized \cite{Berez}. Whereas these types of QD systems are not accessible using optical techniques, self-assembled QDs provide optical access via confined excitonic states and transitions. The atom-like properties of such QDs, for example, single-photon emission \cite{Michl} and single-spins \cite{Brack, Atatu} make them possible candidates for the realization of quantum gates \cite{Bonad,Li} and for quantum information processing and quantum computation \cite{Imamo,Roble} (and refs. therein).
Optical experiments on self-assembled QDs coupled to different kinds of micro- and nano-resonators \cite{Reith,Badol} have been reported as well as artificial molecules created by stacking \cite{Solom,Bayer,Krenn,Ortne,Stina,Krenn2} and laterally arranging QDs \cite{Schmi,Songm,Beirn}. Recently conditional quantum dynamics has been presented in an interacting vertical double-QD system where the quantum state of one dot controls the measurement result in the other dot \cite{Roble}.

For possible applications of QD-systems in quantum information processing both single-photon quality and high optical efficiency are important quantities. 
Signal enhancement of single-photon sources has been reported for single QDs embedded into cavity structures \cite{Vahal}. Also, QDs have been used as a source for polarization entangled photons \cite{Steve, Akopi, Hafen}. In refs. \cite{Steve, Hafen} polarization entanglement has been achieved between biexciton-exciton-pairs necessarily having excitonic fine-structure splittings (FSS) that are small compared to the corresponding radiative life times.

In this work laterally coupled QDs are optically studied with an enhanced luminescence rate achieved by incorporating the QD molecules (QDMs) into a planar micro-cavity. Furthermore, the excitonic structure is investigated in order to resolve possible FSSs which are expected in the lateral QDM due to their lower symmetry as compared to single QDs.

\section{\label{sec:Cavity}Lateral InGaAs QDMs in a lambda-cavity}

\subsection{\label{sec:Cavity_Sample}Sample fabrication with and without cavity}

The QDM samples without a cavity were grown on GaAs$\left(001\right)$ substrates using a unique technique combining molecular beam epitaxy (MBE) and \textit{in situ} etching \cite{Schmi,Songm,Wang}. Using this growth method, samples could be produced that contain mainly bi-molecules, i.e. QDMs that are composed of two individual dots, at structural densities as low as $10^8$ cm$^{-2}$. The average edge-to-edge distance between two QDs forming a bi-molecule is between 6 and 8 nm with individual dot sizes of about 40 nm in diameter and 2.6 nm in height \cite{Wang}. Furthermore, all bi-molecules are oriented along the same crystallographic direction, $\left[1\bar{1}0\right]$, which in turn defines the QDM inter-dot coupling axis \cite{Beirn}. For the optical experiments the QDMs were partially capped and annealed to blue-shift the emission energy to around 1.3 eV to enable the use of Si-based detectors and were overgrown with a 100 nm GaAs cap layer to improve the optical quality. The fact that the QDMs are oriented parallel to one another, with only small deviations of a few percent, allows for a simple means to manipulate the inter-dot coupling. In order to do this parallel chromium/gold Schottky electrode structures with separations between 5 and 30 $\mu$m were placed on the sample surface by photo-lithography to apply homogeneous lateral electric fields along the molecular coupling axis \cite{Beirn}.

\begin{figure}
	\centering
		\includegraphics{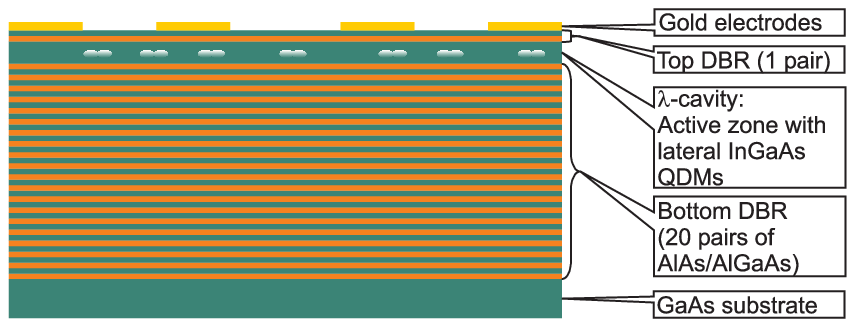}
	\caption{Schematic cross-section of the $\lambda$-cavity sample}
	\label{fig:Fig1}
\end{figure}

To increase the photoluminescence (PL) intensity from the QDMs they were embedded in a planar cavity. The smallest achievable distance between the molecules and electrodes on the surface was used to ensure the homogeneity of the electric field, which was realized by creating a $\lambda$-cavity. The aforementioned growth process has been modified in the following way: in order to do this first, a distributed Bragg reflector (DBR) was grown on a GaAs$\left(001\right)$ substrate using metal-organic vapor phase epitaxy (MOVPE). MOVPE has been chosen for the DBR growth because it offers a significantly higher growth rate compared to MBE and therefore provides a faster and more efficient production process for homogeneous multi-layer structures \cite{Rossb}. The DBR contains 20 $\lambda/4$-periods of AlAs/Al$_{0.2}$Ga$_{0.8}$As with a nominal thickness that corresponds to the average QDM s-shell emission energy of around 1.35 eV. 
The bottom DBR was terminated with a 40 nm GaAs buffer layer in order to avoid oxidation. 
The growth was then continued by MBE. After thermal deoxidation, the bottom part of the $\lambda$-cavity was completed by growing 50 nm GaAs buffer, followed by InAs QDs with high density capped with 40 nm GaAs. The slightly rough GaAs surface produced by buried InAs QDs is used to enhance the yield of QDMs \cite{Wang}. Then InAs QDMs with low density were grown and capped with a $\lambda$/2 thick layer of GaAs. An \textit{in situ} partial cap and annealing step was applied to blue-shift the emission of the QDMs with respect to the buried InAs QDs \cite{Wang}. Finally, a single top AlAs/GaAs pair was grown, completing the cavity structure. 
On top of the epitaxial structure the same type of electrodes was processed as described above. A schematic cross-section of the $\lambda$-cavity sample is shown in Fig.~\ref{fig:Fig1}.

\subsection{\label{sec:Cavity_Lumi}Luminescence enhancement}

\begin{figure}
	\centering
		\includegraphics{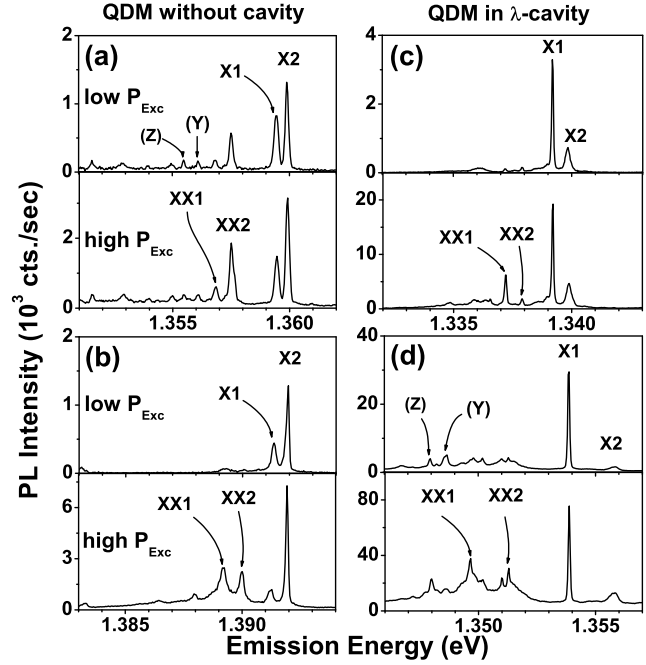}
	\caption{Micro-PL spectra of four different QDMs under non-resonant continuous wave excitation at low power (upper panels) and high power at the exciton saturation level (lower panels): \textbf{(a, b)} QDM without cavity, \textbf{(c)} QDM in a $\lambda$-cavity with a slight mismatch with respect to the center wavelength, \textbf{(d)} QDM in a $\lambda$-cavity that matches the center wavelength.}
	\label{fig:Fig2}
\end{figure}

The micro-PL and photon statistics experiments were performed at 4 K using a confocal microscope setup with a focal spot of about 1 $\mu$m, which, together with the low structural QDM density, allows for single-QDM spectroscopy. A tunable Ti:sapphire continuous wave (cw) laser (700 nm - 1 $\mu$m) was used for optical excitation. The QDM PL was dispersed using a 0.5 m and a 0.75 m spectrometer and collected using either a liquid nitrogen-cooled charge coupled device (CCD) or a Hanbury-Brown and Twiss (HBT) type setup \cite{Hanbu} with two avalanche photodiodes. Depending on the used grating and spectrometer length, the spectral energy resolution was between 10 and 30 $\mu$eV.
A typical s-shell PL-spectrum of a single QDM contains up to six emission lines depending on the charging situation and the excitation power. In the spectra displayed in Fig.~\ref{fig:Fig2} at low excitation power two PL lines, X1 and X2, dominate the spectra which are both due to neutral excitonic recombination. Two weaker PL lines, Y and Z, can eventually also be seen (Figs.~\ref{fig:Fig2}(a and d)), which are due to charged excitonic recombination. At higher excitation power two more PL lines, XX1 and XX2, emerge and increase with a super-linear power dependence indicating that both are due to neutral biexcitonic recombination. A more detailed discussion of the single-QDM PL-spectra and the subsequent identification of their origin can be found in refs.~\cite{Beirn,Herma}.
The maximum QDM photon emission rate along the setup beam axis has been calculated using the measured integrated excitonic saturation count-rates under non-resonant cw excitation by considering the signal transmission through all optical components ($T\approx 40 \%$) and the CCD quantum efficiency ($QE\approx 60 \%$). For a typical QDM without a $\lambda$-cavity the rate is around 20 - 30 kHz (photons/second) (Fig.~\ref{fig:Fig2}(a, b)), where the second QDM is the one that exhibits the strongest luminescence that was observed for some 100 molecules measured so far. The PL emission collection efficiency can be strongly increased when the QDMs are placed in a $\lambda$-cavity. Even for the emission of a QDM with a small mismatch with respect to the cavity center wavelength a maximum emission rate of 200 kHz (Fig.~\ref{fig:Fig2}(c)) is observed which increases further to 850 kHz when the emission matches the cavity mode (Fig.~\ref{fig:Fig2}(d)). 
This 30 times enhancement of the single-QDM PL is mainly attributed to a higher collection efficiency as a result of the emitted photons being preferentially directed along the growth-direction and perpendicular to the DBR-planes.

No coupling effects between the QDMs and the cavity could be observed as expected for a planar cavity structure. This is because of the low quality factor ($Q<100$) and large mode volume $V$ which results in a Purcell factor ($F_P\propto\left(Q/V\right)$) that is very small \cite{Badol}. This could be experimentally seen when comparing the decay times of neutral excitonic recombination from QDMs with and without a $\lambda$-cavity. For both types of samples decay times of around 1 ns were measured for the excitons and 0.5 ns for the biexcitons (discussed in more detail in ref.~\cite{Herma}). Figs.~\ref{fig:Fig3}((a) and (c)) also display decay times of this value extracted from the fits to the autocorrelation measurements of both X1 and X2.

\subsection{\label{sec:Cavity_Efield}Electric field tuning}

\begin{figure}
	\centering
		\includegraphics{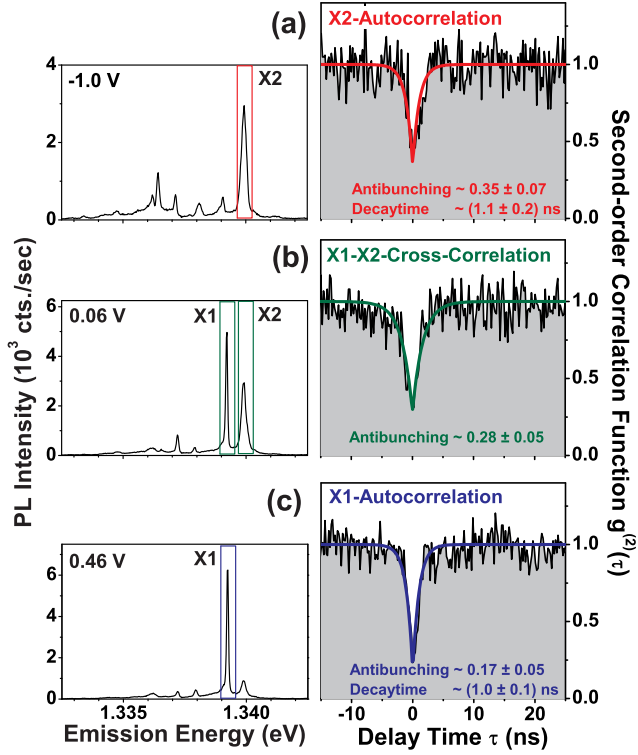}
	\caption{Electric field tuning of the QDM PL under non-resonant cw excitation (1.433 eV, 300 W/cm$^2$): the left panel displays PL spectra recorded at three different voltages, the right panel displays the corresponding intensity correlation measurements. \textbf{(a)} At -1.0 V X2 is the dominant exciton line, the fit to the autocorrelation data gives an antibunching value of 0.35 $\pm$ 0.07 and a decay time of 1.1 $\pm$ 0.2 ns; \textbf{(b)} at 0.06 V X1 and X2 have the same integrated intensity, the fit to their cross-correlation data gives an antibunching value of 0.28 $\pm$ 0.05; \textbf{(c)} at 0.46 V X1 is the dominant exciton line, the fit to the autocorrelation data gives an antibunching value of 0.17 $\pm$ 0.05 and a decay time of 1.0 $\pm$ 0.1 ns.}
	\label{fig:Fig3}
\end{figure}

Electron-tunneling has been proposed to be the dominant coupling mechanism in our lateral QDMs \cite{Beirn}. Using an electric field applied parallel to the QDM coupling axis it is possible to reversibly shift the electron probability density from one dot to the other as manifested by the relative switching between the two neutral exciton recombination lines X1 and X2. In the situation displayed in Fig.~\ref{fig:Fig3}(a), at an applied bias voltage of -1.0 V between the two electrodes that are spaced by 10 $\mu$m, X2 is the dominant PL line. The corresponding measurement of the intensity autocorrelation function of X2 shows clear single-photon emission character with an antibunching value of 0.35. At a voltage of 0.06 V for this particular molecule the two single-dot electron ground state energies are energetically aligned and the X1 and X2 emission lines are equally intense (integrated count-rates). The corresponding intensity cross-correlation measurement between X1 and X2 photons that were used as the \textit{Stop} and \textit{Start} signal in the HBT setup is shown in Fig.~\ref{fig:Fig3}(b). The antibunching value of 0.28 is also well below $0.5$ which is the upper limit for a single quantum emitter. Therefore, both PL lines, X1 and X2, are emitted from the same quantum system, that is, the lateral QDM. When the bias voltage is further increased to 0.46 V, X1 becomes the dominant PL line as shown in Fig.~\ref{fig:Fig3}(c). Again the corresponding intensity autocorrelation function displays single-photon emission character with an antibunching value of 0.17.

\section{\label{sec:Pol}Polarization Fine-structure}

Studying the excitonic polarization fine-structure is of particular interest for two reasons as it allows one, firstly, to identify a biexciton-exciton cascade and, secondly, to determine the exciton FSS. The relative value of the FSS of the exciton state ($\Delta E_{FS}$) with respect to the corresponding natural radiative line width ($\Delta E_{rad}$) is decisive in terms of the correlation character of the biexciton-exciton pair (Fig.~1 in \cite{Hafen}). For $\Delta E_{FS}/\Delta E_{rad}>1$, the two orthogonally polarized exciton-biexciton photon pairs exhibit a classical polarization correlation, whereas for $\Delta E_{FS}/\Delta E_{rad}<1$ they are polarization entangled to a certain extent. It has recently been reported that the polarization FSS of single QD excitons can be tuned using external magnetic and electric fields \cite{Bayer2,Vogel}. Therefore, for an initial FSS that is sufficiently small tuning down to values close to zero can be achieved and so the QD under investigation represents a potential source of polarization entangled photon pairs.

\begin{figure}
	\centering
		\includegraphics{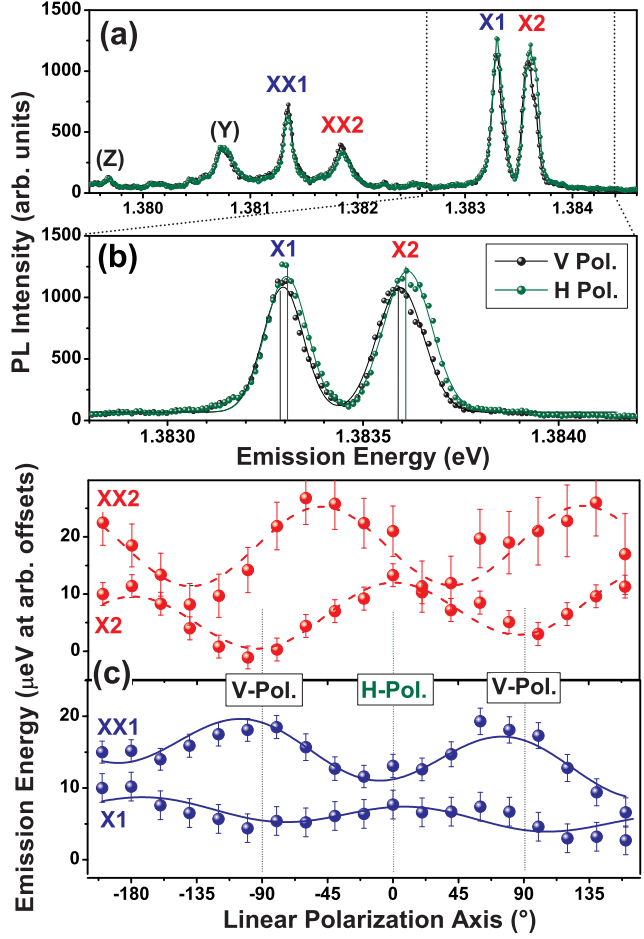}
	\caption{ \textbf{(a)} Polarization-resolved QDM PL showing the complete s-shell spectrum for the two orthogonal linear polarizations \emph{H} and \emph{V}; \textbf{(b)} a smaller range showing the neutral excitons X1 and X2 in which the dots are the measurement data and the solid lines are fits to the data; \textbf{(c)} the fitted emission peak positions for the two radiative cascades as a function of linear polarization axis are plotted at arbitrary offsets to illustrate their polarization correlation: the lower panel shows the first cascade (XX1-X1), the upper panel shows the second cascade (XX2-X2).}
	\label{fig:Fig4}
\end{figure}

The highest available spectral energy resolution of around 10 $\mu$eV (using a 1800 lines/mm grating and the 0.75 m spectrometer) was used to resolve a possible FSS of the excitonic emission from a QDM without a cavity. Micro-PL spectra have been recorded as a function of linear polarization by using a rotatable $\lambda /2$-plate and a fixed polarizing cube beam splitter. The \emph{H}-axis corresponds to the molecular coupling axis and has been defined as the zero linear polarization angle. Fig.~\ref{fig:Fig4}(a) shows the complete s-shell emission of a QDM for the two orthogonal linear polarizations \emph{H} and \emph{V}. At line widths of around 150 $\mu$eV a spectral offset on the order of 10 $\mu$eV between the neutral excitons can be seen when comparing the smaller range spectra in Fig.~\ref{fig:Fig4}(b) that were taken at \emph{H}- and \emph{V}-polarization. For the PL lines Y and Z, which have been assigned to charged exciton transitions earlier no such splitting appears to be present as expected \cite{Bayer2,Vogel}. 

A detailed analysis has been carried out on the two biexciton-exciton cascades: XX1-X1 (blue, solid lines in lower panel of Fig.~\ref{fig:Fig4}(c)) and XX2-X2 (red, dashed lines in upper panel of Fig.~\ref{fig:Fig4}(c)). An oscillation of the emission energies can be seen for both excitons and biexcitons with a period of 180$^0$. This is indicative of a linear polarization splitting on the order of the difference between the minima and maxima of the emission energies which averages to $\Delta E_{FS}\approx 10 \mu$eV. As expected from single-QD results both cascades under investigation show the characteristic behavior where the low energy biexciton component is followed by the high energy exciton component and vice versa. For example at a polarization angle around \emph{H} the low energy XX1 (XX2) component is followed by the high energy X1 (X2) component, whereas at a polarization angle around \emph{V} the high energy XX1 (XX2) component is followed by the low energy X1 (X2) component. The offset of around +35$^0$ in the extremal emission energies for XX2 with respect to the other three emission lines may be the result of a contribution of differently polarized PL background. XX2 is the most susceptible to background contribution since it exhibits the weakest count-rate of all analyzed PL lines. The observed polarization behavior supports the assignment of XX1-X1 and XX2-X2 as the two biexciton-exciton cascades which have been reported earlier via photon cross-correlation measurements between XX1 and X1. A detailed discussion can be found in ref.~\cite{Beirn} in which the second-order correlation function shows antibunching for $\tau<0$ (emission from one single-quantum system) and bunching for $\tau>0$ (enhanced probability of X1 emission after XX1 emission in a radiative cascade).

\section{\label{sec:Sum}Conclusion and Outlook}

Using a novel approach to epitaxially produce QDMs in a planar cavity a 30 times enhancement in luminescence could be observed. The presented high optical quality of the QDM single-photon emission together with the strongly enhanced luminescence are essential requirements for further photon statistics experiments and time-resolved studies of such structures.
The ongoing work will focus on the charge carrier dynamics and the polarization properties of the coupled system, as well as on the excited state structure.

With a rather small FSS of around $10 \mu eV$ found for the molecular excitons it should be possible to tune it to zero. As one promising approach we plan to use a two-dimensional geometry of electrodes to apply lateral electric fields at arbitrary directions with respect to the QDM coupling axis. in this way the FSS could be tuned to zero and the emission of the two excitons could also be switched. This would enable us to use the QDM as a switchable source of two pairs of entangled photons from the same quantum system.

\begin{acknowledgments}
The authors gratefully acknowledge financial support from the Deutsche Forschungsgemeinschaft via grants FOR 730 and SFB/TR 21.
\end{acknowledgments}




\newpage

\newpage

\newpage

\newpage

\newpage

\begin{thebibliography}{33}
\expandafter\ifx\csname natexlab\endcsname\relax\def\natexlab#1{#1}\fi
\expandafter\ifx\csname bibnamefont\endcsname\relax
  \def\bibnamefont#1{#1}\fi
\expandafter\ifx\csname bibfnamefont\endcsname\relax
  \def\bibfnamefont#1{#1}\fi
\expandafter\ifx\csname citenamefont\endcsname\relax
  \def\citenamefont#1{#1}\fi
\expandafter\ifx\csname url\endcsname\relax
  \def\url#1{\texttt{#1}}\fi
\expandafter\ifx\csname urlprefix\endcsname\relax\def\urlprefix{URL }\fi
\providecommand{\bibinfo}[2]{#2}
\providecommand{\eprint}[2][]{\url{#2}}

\bibitem[{\citenamefont{Hayashi et~al.}(2003)\citenamefont{Hayashi, Fujisawa,
  Cheong, Jeong, and Hirayama}}]{Hayas}
\bibinfo{author}{\bibfnamefont{T.}~\bibnamefont{Hayashi}},
  \bibinfo{author}{\bibfnamefont{T.}~\bibnamefont{Fujisawa}},
  \bibinfo{author}{\bibfnamefont{H.~D.} \bibnamefont{Cheong}},
  \bibinfo{author}{\bibfnamefont{Y.~H.} \bibnamefont{Jeong}}, \bibnamefont{and}
  \bibinfo{author}{\bibfnamefont{Y.}~\bibnamefont{Hirayama}},
  \bibinfo{journal}{Phys.\ Rev.\ Lett.} \textbf{\bibinfo{volume}{91}},
  \bibinfo{pages}{226804} (\bibinfo{year}{2003}).

\bibitem[{\citenamefont{Petta et~al.}(2005)\citenamefont{Petta, Johnson,
  Taylor, Laird, Yacoby, Lukin, Marcus, Hanson, and Gossard}}]{Petta}
\bibinfo{author}{\bibfnamefont{J.~R.} \bibnamefont{Petta}},
  \bibinfo{author}{\bibfnamefont{A.~C.} \bibnamefont{Johnson}},
  \bibinfo{author}{\bibfnamefont{J.~M.} \bibnamefont{Taylor}},
  \bibinfo{author}{\bibfnamefont{E.~A.} \bibnamefont{Laird}},
  \bibinfo{author}{\bibfnamefont{A.}~\bibnamefont{Yacoby}},
  \bibinfo{author}{\bibfnamefont{M.~D.} \bibnamefont{Lukin}},
  \bibinfo{author}{\bibfnamefont{C.~M.} \bibnamefont{Marcus}},
  \bibinfo{author}{\bibfnamefont{M.~P.} \bibnamefont{Hanson}},
  \bibnamefont{and} \bibinfo{author}{\bibfnamefont{A.~C.}
  \bibnamefont{Gossard}}, \bibinfo{journal}{Science}
  \textbf{\bibinfo{volume}{309}}, \bibinfo{pages}{2180} (\bibinfo{year}{2005}).

\bibitem[{\citenamefont{Koppens et~al.}(2006)\citenamefont{Koppens, Buizert,
  Tielrooij, Vink, Nowack, Meunier, Kouwenhoven, and Vandersypen}}]{Koppe}
\bibinfo{author}{\bibfnamefont{F.~H.~L.} \bibnamefont{Koppens}},
  \bibinfo{author}{\bibfnamefont{C.}~\bibnamefont{Buizert}},
  \bibinfo{author}{\bibfnamefont{K.~J.} \bibnamefont{Tielrooij}},
  \bibinfo{author}{\bibfnamefont{I.~T.} \bibnamefont{Vink}},
  \bibinfo{author}{\bibfnamefont{K.~C.} \bibnamefont{Nowack}},
  \bibinfo{author}{\bibfnamefont{T.}~\bibnamefont{Meunier}},
  \bibinfo{author}{\bibfnamefont{L.~P.} \bibnamefont{Kouwenhoven}},
  \bibnamefont{and} \bibinfo{author}{\bibfnamefont{L.~M.~K.}
  \bibnamefont{Vandersypen}}, \bibinfo{journal}{Nature}
  \textbf{\bibinfo{volume}{442}}, \bibinfo{pages}{766} (\bibinfo{year}{2006}).

\bibitem[{\citenamefont{Elzerman et~al.}(2004)\citenamefont{Elzerman, Hanson,
  Willems~van Beveren, Witkamp, Vandersypen, and Kouwenhoven}}]{Elzer}
\bibinfo{author}{\bibfnamefont{J.~M.} \bibnamefont{Elzerman}},
  \bibinfo{author}{\bibfnamefont{R.}~\bibnamefont{Hanson}},
  \bibinfo{author}{\bibfnamefont{L.~H.} \bibnamefont{Willems~van Beveren}},
  \bibinfo{author}{\bibfnamefont{B.}~\bibnamefont{Witkamp}},
  \bibinfo{author}{\bibfnamefont{L.~M.~K.} \bibnamefont{Vandersypen}},
  \bibnamefont{and} \bibinfo{author}{\bibfnamefont{L.~P.}
  \bibnamefont{Kouwenhoven}}, \bibinfo{journal}{Nature}
  \textbf{\bibinfo{volume}{430}}, \bibinfo{pages}{431} (\bibinfo{year}{2004}).

\bibitem[{\citenamefont{Berezovsky et~al.}(2008)\citenamefont{Berezovsky,
  Mikkelsen, Stoltz, Coldren, and Awschalom}}]{Berez}
\bibinfo{author}{\bibfnamefont{J.}~\bibnamefont{Berezovsky}},
  \bibinfo{author}{\bibfnamefont{M.~H.} \bibnamefont{Mikkelsen}},
  \bibinfo{author}{\bibfnamefont{N.~G.} \bibnamefont{Stoltz}},
  \bibinfo{author}{\bibfnamefont{L.~A.} \bibnamefont{Coldren}},
  \bibnamefont{and} \bibinfo{author}{\bibfnamefont{D.~D.}
  \bibnamefont{Awschalom}}, \bibinfo{journal}{Science}
  \textbf{\bibinfo{volume}{320}}, \bibinfo{pages}{349} (\bibinfo{year}{2008}).

\bibitem[{\citenamefont{Michler et~al.}(2000)\citenamefont{Michler, Kiraz,
  Becher, Schoenfeld, Petroff, Zhang, Hu, and Imamoglu}}]{Michl}
\bibinfo{author}{\bibfnamefont{P.}~\bibnamefont{Michler}},
  \bibinfo{author}{\bibfnamefont{A.}~\bibnamefont{Kiraz}},
  \bibinfo{author}{\bibfnamefont{C.}~\bibnamefont{Becher}},
  \bibinfo{author}{\bibfnamefont{W.~V.} \bibnamefont{Schoenfeld}},
  \bibinfo{author}{\bibfnamefont{P.~M.} \bibnamefont{Petroff}},
  \bibinfo{author}{\bibfnamefont{L.~D.} \bibnamefont{Zhang}},
  \bibinfo{author}{\bibfnamefont{E.}~\bibnamefont{Hu}}, \bibnamefont{and}
  \bibinfo{author}{\bibfnamefont{A.}~\bibnamefont{Imamoglu}},
  \bibinfo{journal}{Science} \textbf{\bibinfo{volume}{290}},
  \bibinfo{pages}{2282} (\bibinfo{year}{2000}).

\bibitem[{\citenamefont{Bracker et~al.}(2005)\citenamefont{Bracker, Stinaff,
  Gammon, Ware, Tischler, Shabaev, Efros, Park, Gershoni, Korenev
  et~al.}}]{Brack}
\bibinfo{author}{\bibfnamefont{A.~S.} \bibnamefont{Bracker}},
  \bibinfo{author}{\bibfnamefont{E.~A.} \bibnamefont{Stinaff}},
  \bibinfo{author}{\bibfnamefont{D.}~\bibnamefont{Gammon}},
  \bibinfo{author}{\bibfnamefont{M.~E.} \bibnamefont{Ware}},
  \bibinfo{author}{\bibfnamefont{J.~G.} \bibnamefont{Tischler}},
  \bibinfo{author}{\bibfnamefont{A.}~\bibnamefont{Shabaev}},
  \bibinfo{author}{\bibfnamefont{A.~L.} \bibnamefont{Efros}},
  \bibinfo{author}{\bibfnamefont{D.}~\bibnamefont{Park}},
  \bibinfo{author}{\bibfnamefont{D.}~\bibnamefont{Gershoni}},
  \bibinfo{author}{\bibfnamefont{V.~L.} \bibnamefont{Korenev}},
  \bibnamefont{et~al.}, \bibinfo{journal}{Phys.\ Rev.\ Lett.}
  \textbf{\bibinfo{volume}{94}}, \bibinfo{pages}{047402}
  (\bibinfo{year}{2005}).

\bibitem[{\citenamefont{Atature et~al.}(2006)\citenamefont{Atature, Dreiser,
  Badolato, Hogele, Karrai, and Imamoglu}}]{Atatu}
\bibinfo{author}{\bibfnamefont{M.}~\bibnamefont{Atature}},
  \bibinfo{author}{\bibfnamefont{J.}~\bibnamefont{Dreiser}},
  \bibinfo{author}{\bibfnamefont{A.}~\bibnamefont{Badolato}},
  \bibinfo{author}{\bibfnamefont{A.}~\bibnamefont{Hogele}},
  \bibinfo{author}{\bibfnamefont{K.}~\bibnamefont{Karrai}}, \bibnamefont{and}
  \bibinfo{author}{\bibfnamefont{A.}~\bibnamefont{Imamoglu}},
  \bibinfo{journal}{Science} \textbf{\bibinfo{volume}{312}},
  \bibinfo{pages}{551} (\bibinfo{year}{2006}).

\bibitem[{\citenamefont{Bonadeo et~al.}(1998)\citenamefont{Bonadeo, Erland,
  Gammon, Park, Katzer, and Steel}}]{Bonad}
\bibinfo{author}{\bibfnamefont{N.~H.} \bibnamefont{Bonadeo}},
  \bibinfo{author}{\bibfnamefont{J.}~\bibnamefont{Erland}},
  \bibinfo{author}{\bibfnamefont{D.}~\bibnamefont{Gammon}},
  \bibinfo{author}{\bibfnamefont{D.}~\bibnamefont{Park}},
  \bibinfo{author}{\bibfnamefont{D.~S.} \bibnamefont{Katzer}},
  \bibnamefont{and} \bibinfo{author}{\bibfnamefont{D.~G.} \bibnamefont{Steel}},
  \bibinfo{journal}{Science} \textbf{\bibinfo{volume}{282}},
  \bibinfo{pages}{1473} (\bibinfo{year}{1998}).

\bibitem[{\citenamefont{Li et~al.}(2003)\citenamefont{Li, Wu, Steel, Gammon,
  Stievater, Katzer, Park, Piermarocchi, and Sham}}]{Li}
\bibinfo{author}{\bibfnamefont{X.~Q.} \bibnamefont{Li}},
  \bibinfo{author}{\bibfnamefont{Y.~W.} \bibnamefont{Wu}},
  \bibinfo{author}{\bibfnamefont{D.}~\bibnamefont{Steel}},
  \bibinfo{author}{\bibfnamefont{D.}~\bibnamefont{Gammon}},
  \bibinfo{author}{\bibfnamefont{T.~H.} \bibnamefont{Stievater}},
  \bibinfo{author}{\bibfnamefont{D.~S.} \bibnamefont{Katzer}},
  \bibinfo{author}{\bibfnamefont{D.}~\bibnamefont{Park}},
  \bibinfo{author}{\bibfnamefont{C.}~\bibnamefont{Piermarocchi}},
  \bibnamefont{and} \bibinfo{author}{\bibfnamefont{L.~J.} \bibnamefont{Sham}},
  \bibinfo{journal}{Science} \textbf{\bibinfo{volume}{301}},
  \bibinfo{pages}{809} (\bibinfo{year}{2003}).

\bibitem[{\citenamefont{Imamoglu et~al.}(1999)\citenamefont{Imamoglu,
  Awschalom, Burkard, DiVincenzo, Loss, Sherwin, and Small}}]{Imamo}
\bibinfo{author}{\bibfnamefont{A.}~\bibnamefont{Imamoglu}},
  \bibinfo{author}{\bibfnamefont{D.}~\bibnamefont{Awschalom}},
  \bibinfo{author}{\bibfnamefont{G.}~\bibnamefont{Burkard}},
  \bibinfo{author}{\bibfnamefont{D.}~\bibnamefont{DiVincenzo}},
  \bibinfo{author}{\bibfnamefont{D.}~\bibnamefont{Loss}},
  \bibinfo{author}{\bibfnamefont{M.}~\bibnamefont{Sherwin}}, \bibnamefont{and}
  \bibinfo{author}{\bibfnamefont{A.}~\bibnamefont{Small}},
  \bibinfo{journal}{Phys.\ Rev.\ Lett.} \textbf{\bibinfo{volume}{83}},
  \bibinfo{pages}{4204} (\bibinfo{year}{1999}).

\bibitem[{\citenamefont{Robledo et~al.}(2008)\citenamefont{Robledo, Elzerman,
  Jundt, Atature, Hogele, Falt, and Imamoglu}}]{Roble}
\bibinfo{author}{\bibfnamefont{L.}~\bibnamefont{Robledo}},
  \bibinfo{author}{\bibfnamefont{J.}~\bibnamefont{Elzerman}},
  \bibinfo{author}{\bibfnamefont{G.}~\bibnamefont{Jundt}},
  \bibinfo{author}{\bibfnamefont{M.}~\bibnamefont{Atature}},
  \bibinfo{author}{\bibfnamefont{A.}~\bibnamefont{Hogele}},
  \bibinfo{author}{\bibfnamefont{S.}~\bibnamefont{Falt}}, \bibnamefont{and}
  \bibinfo{author}{\bibfnamefont{A.}~\bibnamefont{Imamoglu}},
  \bibinfo{journal}{Science} \textbf{\bibinfo{volume}{320}},
  \bibinfo{pages}{772} (\bibinfo{year}{2008}).

\bibitem[{\citenamefont{Reithmaier et~al.}(2004)\citenamefont{Reithmaier, Sek,
  Loffler, Hofmann, Kuhn, Reitzenstein, Keldysh, Kulakovskii, Reinecke, and
  Forchel}}]{Reith}
\bibinfo{author}{\bibfnamefont{J.}~\bibnamefont{Reithmaier}},
  \bibinfo{author}{\bibfnamefont{G.}~\bibnamefont{Sek}},
  \bibinfo{author}{\bibfnamefont{A.}~\bibnamefont{Loffler}},
  \bibinfo{author}{\bibfnamefont{C.}~\bibnamefont{Hofmann}},
  \bibinfo{author}{\bibfnamefont{S.}~\bibnamefont{Kuhn}},
  \bibinfo{author}{\bibfnamefont{S.}~\bibnamefont{Reitzenstein}},
  \bibinfo{author}{\bibfnamefont{L.}~\bibnamefont{Keldysh}},
  \bibinfo{author}{\bibfnamefont{V.}~\bibnamefont{Kulakovskii}},
  \bibinfo{author}{\bibfnamefont{T.}~\bibnamefont{Reinecke}}, \bibnamefont{and}
  \bibinfo{author}{\bibfnamefont{A.}~\bibnamefont{Forchel}},
  \bibinfo{journal}{Nature} \textbf{\bibinfo{volume}{432}},
  \bibinfo{pages}{197} (\bibinfo{year}{2004}).

\bibitem[{\citenamefont{Badolato et~al.}(2005)\citenamefont{Badolato, Hennessy,
  Atature, Dreiser, Hu, Petroff, and Imamoglu}}]{Badol}
\bibinfo{author}{\bibfnamefont{A.}~\bibnamefont{Badolato}},
  \bibinfo{author}{\bibfnamefont{K.}~\bibnamefont{Hennessy}},
  \bibinfo{author}{\bibfnamefont{M.}~\bibnamefont{Atature}},
  \bibinfo{author}{\bibfnamefont{J.}~\bibnamefont{Dreiser}},
  \bibinfo{author}{\bibfnamefont{E.}~\bibnamefont{Hu}},
  \bibinfo{author}{\bibfnamefont{P.}~\bibnamefont{Petroff}}, \bibnamefont{and}
  \bibinfo{author}{\bibfnamefont{A.}~\bibnamefont{Imamoglu}},
  \bibinfo{journal}{Science} \textbf{\bibinfo{volume}{308}},
  \bibinfo{pages}{1158} (\bibinfo{year}{2005}).

\bibitem[{\citenamefont{Solomon et~al.}(1996)\citenamefont{Solomon, Trezza,
  Marshall, and Harris}}]{Solom}
\bibinfo{author}{\bibfnamefont{G.}~\bibnamefont{Solomon}},
  \bibinfo{author}{\bibfnamefont{J.}~\bibnamefont{Trezza}},
  \bibinfo{author}{\bibfnamefont{A.}~\bibnamefont{Marshall}}, \bibnamefont{and}
  \bibinfo{author}{\bibfnamefont{J.}~\bibnamefont{Harris}},
  \bibinfo{journal}{Phys.\ Rev.\ Lett.} \textbf{\bibinfo{volume}{76}},
  \bibinfo{pages}{952} (\bibinfo{year}{1996}).

\bibitem[{\citenamefont{Bayer et~al.}(2001)\citenamefont{Bayer, Hawrylak,
  Hinzer, Fafard, Korkusinski, Wasilewski, Stern, and Forchel}}]{Bayer}
\bibinfo{author}{\bibfnamefont{M.}~\bibnamefont{Bayer}},
  \bibinfo{author}{\bibfnamefont{P.}~\bibnamefont{Hawrylak}},
  \bibinfo{author}{\bibfnamefont{K.}~\bibnamefont{Hinzer}},
  \bibinfo{author}{\bibfnamefont{S.}~\bibnamefont{Fafard}},
  \bibinfo{author}{\bibfnamefont{M.}~\bibnamefont{Korkusinski}},
  \bibinfo{author}{\bibfnamefont{Z.~R.} \bibnamefont{Wasilewski}},
  \bibinfo{author}{\bibfnamefont{O.}~\bibnamefont{Stern}}, \bibnamefont{and}
  \bibinfo{author}{\bibfnamefont{A.}~\bibnamefont{Forchel}},
  \bibinfo{journal}{Science} \textbf{\bibinfo{volume}{291}},
  \bibinfo{pages}{451} (\bibinfo{year}{2001}).

\bibitem[{\citenamefont{Krenner et~al.}(2005)\citenamefont{Krenner, Sabathil,
  Clark, Kress, Schuh, Bichler, Abstreiter, and Finley}}]{Krenn}
\bibinfo{author}{\bibfnamefont{H.~J.} \bibnamefont{Krenner}},
  \bibinfo{author}{\bibfnamefont{M.}~\bibnamefont{Sabathil}},
  \bibinfo{author}{\bibfnamefont{E.~C.} \bibnamefont{Clark}},
  \bibinfo{author}{\bibfnamefont{A.}~\bibnamefont{Kress}},
  \bibinfo{author}{\bibfnamefont{D.}~\bibnamefont{Schuh}},
  \bibinfo{author}{\bibfnamefont{M.}~\bibnamefont{Bichler}},
  \bibinfo{author}{\bibfnamefont{G.}~\bibnamefont{Abstreiter}},
  \bibnamefont{and} \bibinfo{author}{\bibfnamefont{J.~J.}
  \bibnamefont{Finley}}, \bibinfo{journal}{Phys.\ Rev.\ Lett.}
  \textbf{\bibinfo{volume}{94}}, \bibinfo{pages}{057402}
  (\bibinfo{year}{2005}).

\bibitem[{\citenamefont{Ortner et~al.}(2005)\citenamefont{Ortner, Bayer,
  Lyanda-Geller, Reinecke, Kress, Reithmaier, and Forchel}}]{Ortne}
\bibinfo{author}{\bibfnamefont{G.}~\bibnamefont{Ortner}},
  \bibinfo{author}{\bibfnamefont{M.}~\bibnamefont{Bayer}},
  \bibinfo{author}{\bibfnamefont{Y.}~\bibnamefont{Lyanda-Geller}},
  \bibinfo{author}{\bibfnamefont{T.~L.} \bibnamefont{Reinecke}},
  \bibinfo{author}{\bibfnamefont{A.}~\bibnamefont{Kress}},
  \bibinfo{author}{\bibfnamefont{J.~P.} \bibnamefont{Reithmaier}},
  \bibnamefont{and} \bibinfo{author}{\bibfnamefont{A.}~\bibnamefont{Forchel}},
  \bibinfo{journal}{Phys.\ Rev.\ Lett.} \textbf{\bibinfo{volume}{94}},
  \bibinfo{pages}{157401} (\bibinfo{year}{2005}).

\bibitem[{\citenamefont{Stinaff et~al.}(2006)\citenamefont{Stinaff, Scheibner,
  Bracker, Ponomarev, Korenev, Ware, Doty, Reinecke, and Gammon}}]{Stina}
\bibinfo{author}{\bibfnamefont{E.~A.} \bibnamefont{Stinaff}},
  \bibinfo{author}{\bibfnamefont{M.}~\bibnamefont{Scheibner}},
  \bibinfo{author}{\bibfnamefont{A.~S.} \bibnamefont{Bracker}},
  \bibinfo{author}{\bibfnamefont{I.~V.} \bibnamefont{Ponomarev}},
  \bibinfo{author}{\bibfnamefont{V.~L.} \bibnamefont{Korenev}},
  \bibinfo{author}{\bibfnamefont{M.~E.} \bibnamefont{Ware}},
  \bibinfo{author}{\bibfnamefont{M.~F.} \bibnamefont{Doty}},
  \bibinfo{author}{\bibfnamefont{T.~L.} \bibnamefont{Reinecke}},
  \bibnamefont{and} \bibinfo{author}{\bibfnamefont{D.}~\bibnamefont{Gammon}},
  \bibinfo{journal}{Science} \textbf{\bibinfo{volume}{311}},
  \bibinfo{pages}{636} (\bibinfo{year}{2006}).

\bibitem[{\citenamefont{Krenner et~al.}(2006)\citenamefont{Krenner, Clark,
  Nakaoka, Bichler, Scheurer, Abstreiter, and Finley}}]{Krenn2}
\bibinfo{author}{\bibfnamefont{H.~J.} \bibnamefont{Krenner}},
  \bibinfo{author}{\bibfnamefont{E.~C.} \bibnamefont{Clark}},
  \bibinfo{author}{\bibfnamefont{T.}~\bibnamefont{Nakaoka}},
  \bibinfo{author}{\bibfnamefont{M.}~\bibnamefont{Bichler}},
  \bibinfo{author}{\bibfnamefont{C.}~\bibnamefont{Scheurer}},
  \bibinfo{author}{\bibfnamefont{G.}~\bibnamefont{Abstreiter}},
  \bibnamefont{and} \bibinfo{author}{\bibfnamefont{J.~J.}
  \bibnamefont{Finley}}, \bibinfo{journal}{Phys.\ Rev.\ Lett.}
  \textbf{\bibinfo{volume}{97}}, \bibinfo{pages}{076403}
  (\bibinfo{year}{2006}).

\bibitem[{\citenamefont{Schmidt et~al.}(2002)\citenamefont{Schmidt, Deneke,
  Kiravittaya, Songmuang, Heidemeyer, Nakamura, Zapf-Gottwick, M{\"u}ller, and
  Jin-Phillipp}}]{Schmi}
\bibinfo{author}{\bibfnamefont{O.~G.} \bibnamefont{Schmidt}},
  \bibinfo{author}{\bibfnamefont{C.}~\bibnamefont{Deneke}},
  \bibinfo{author}{\bibfnamefont{S.}~\bibnamefont{Kiravittaya}},
  \bibinfo{author}{\bibfnamefont{R.}~\bibnamefont{Songmuang}},
  \bibinfo{author}{\bibfnamefont{H.}~\bibnamefont{Heidemeyer}},
  \bibinfo{author}{\bibfnamefont{Y.}~\bibnamefont{Nakamura}},
  \bibinfo{author}{\bibfnamefont{R.}~\bibnamefont{Zapf-Gottwick}},
  \bibinfo{author}{\bibfnamefont{C.}~\bibnamefont{M{\"u}ller}},
  \bibnamefont{and} \bibinfo{author}{\bibfnamefont{N.~Y.}
  \bibnamefont{Jin-Phillipp}}, \bibinfo{journal}{IEEE\ J.\ Sel.\ Top.\ Quantum\
  Electron.} \textbf{\bibinfo{volume}{8}}, \bibinfo{pages}{1025–34}
  (\bibinfo{year}{2002}).

\bibitem[{\citenamefont{Songmuang et~al.}(2003)\citenamefont{Songmuang,
  Kiravittaya, S., and Schmidt}}]{Songm}
\bibinfo{author}{\bibfnamefont{R.}~\bibnamefont{Songmuang}},
  \bibinfo{author}{\bibnamefont{Kiravittaya}},
  \bibinfo{author}{\bibnamefont{S.}}, \bibnamefont{and}
  \bibinfo{author}{\bibfnamefont{O.~G.} \bibnamefont{Schmidt}},
  \bibinfo{journal}{Appl.\ Phys.\ Lett.} \textbf{\bibinfo{volume}{82}},
  \bibinfo{pages}{2892–4} (\bibinfo{year}{2003}).

\bibitem[{\citenamefont{Beirne et~al.}(2006)\citenamefont{Beirne,
  Hermannst{\"a}dter, Wang, Rastelli, Schmidt, and Michler}}]{Beirn}
\bibinfo{author}{\bibfnamefont{G.~J.} \bibnamefont{Beirne}},
  \bibinfo{author}{\bibfnamefont{C.}~\bibnamefont{Hermannst{\"a}dter}},
  \bibinfo{author}{\bibfnamefont{L.}~\bibnamefont{Wang}},
  \bibinfo{author}{\bibfnamefont{A.}~\bibnamefont{Rastelli}},
  \bibinfo{author}{\bibfnamefont{O.~G.} \bibnamefont{Schmidt}},
  \bibnamefont{and} \bibinfo{author}{\bibfnamefont{P.}~\bibnamefont{Michler}},
  \bibinfo{journal}{Phys.\ Rev.\ Lett.} \textbf{\bibinfo{volume}{96}},
  \bibinfo{pages}{137401} (\bibinfo{year}{2006}).

\bibitem[{\citenamefont{Vahala}(2003)}]{Vahal}
\bibinfo{author}{\bibfnamefont{K.~J.} \bibnamefont{Vahala}},
  \bibinfo{journal}{Nature} \textbf{\bibinfo{volume}{424}},
  \bibinfo{pages}{839} (\bibinfo{year}{2003}).

\bibitem[{\citenamefont{Stevenson et~al.}(2006)\citenamefont{Stevenson, Young,
  Atkinson, Cooper, Ritchie, and Shields}}]{Steve}
\bibinfo{author}{\bibfnamefont{R.~M.} \bibnamefont{Stevenson}},
  \bibinfo{author}{\bibfnamefont{R.~J.} \bibnamefont{Young}},
  \bibinfo{author}{\bibfnamefont{P.}~\bibnamefont{Atkinson}},
  \bibinfo{author}{\bibfnamefont{K.}~\bibnamefont{Cooper}},
  \bibinfo{author}{\bibfnamefont{D.~A.} \bibnamefont{Ritchie}},
  \bibnamefont{and} \bibinfo{author}{\bibfnamefont{A.~J.}
  \bibnamefont{Shields}}, \bibinfo{journal}{Nature}
  \textbf{\bibinfo{volume}{439}}, \bibinfo{pages}{179–82}
  (\bibinfo{year}{2006}).

\bibitem[{\citenamefont{Akopian et~al.}(2006)\citenamefont{Akopian, Lindner,
  Poem, Berlatzky, Avron, and Gershoni}}]{Akopi}
\bibinfo{author}{\bibfnamefont{N.}~\bibnamefont{Akopian}},
  \bibinfo{author}{\bibfnamefont{N.~H.} \bibnamefont{Lindner}},
  \bibinfo{author}{\bibfnamefont{E.}~\bibnamefont{Poem}},
  \bibinfo{author}{\bibfnamefont{Y.}~\bibnamefont{Berlatzky}},
  \bibinfo{author}{\bibfnamefont{J.}~\bibnamefont{Avron}}, \bibnamefont{and}
  \bibinfo{author}{\bibfnamefont{D.}~\bibnamefont{Gershoni}},
  \bibinfo{journal}{Phys.\ Rev.\ Lett.} \textbf{\bibinfo{volume}{96}},
  \bibinfo{pages}{130501} (\bibinfo{year}{2006}).

\bibitem[{\citenamefont{Hafenbrak et~al.}(2007)\citenamefont{Hafenbrak, Ulrich,
  Michler, Wang, Rastelli, and Schmidt}}]{Hafen}
\bibinfo{author}{\bibfnamefont{R.}~\bibnamefont{Hafenbrak}},
  \bibinfo{author}{\bibfnamefont{S.~M.} \bibnamefont{Ulrich}},
  \bibinfo{author}{\bibfnamefont{P.}~\bibnamefont{Michler}},
  \bibinfo{author}{\bibfnamefont{L.}~\bibnamefont{Wang}},
  \bibinfo{author}{\bibfnamefont{A.}~\bibnamefont{Rastelli}}, \bibnamefont{and}
  \bibinfo{author}{\bibfnamefont{O.~G.} \bibnamefont{Schmidt}},
  \bibinfo{journal}{New\ J.\ Phys.} \textbf{\bibinfo{volume}{9}},
  \bibinfo{pages}{315} (\bibinfo{year}{2007}).

\bibitem[{\citenamefont{Wang et~al.}(2008)\citenamefont{Wang, Rastelli,
  Kiravittaya, Atkinson, Ding, Bufon, Hermannst{\"a}dter, Witzany, Beirne,
  Michler et~al.}}]{Wang}
\bibinfo{author}{\bibfnamefont{L.}~\bibnamefont{Wang}},
  \bibinfo{author}{\bibfnamefont{A.}~\bibnamefont{Rastelli}},
  \bibinfo{author}{\bibfnamefont{S.}~\bibnamefont{Kiravittaya}},
  \bibinfo{author}{\bibfnamefont{P.}~\bibnamefont{Atkinson}},
  \bibinfo{author}{\bibfnamefont{F.}~\bibnamefont{Ding}},
  \bibinfo{author}{\bibfnamefont{C.~C.~B.} \bibnamefont{Bufon}},
  \bibinfo{author}{\bibfnamefont{C.}~\bibnamefont{Hermannst{\"a}dter}},
  \bibinfo{author}{\bibfnamefont{M.}~\bibnamefont{Witzany}},
  \bibinfo{author}{\bibfnamefont{G.~J.} \bibnamefont{Beirne}},
  \bibinfo{author}{\bibfnamefont{P.}~\bibnamefont{Michler}},
  \bibnamefont{et~al.}, \bibinfo{journal}{New\ J.\ Phys.}
  \textbf{\bibinfo{volume}{10}}, \bibinfo{pages}{045010}
  (\bibinfo{year}{2008}).

\bibitem[{\citenamefont{Roßbach et~al.}(2008)\citenamefont{Roßbach, Schulz,
  Reischle, Beirne, Jetter, and Michler}}]{Rossb}
\bibinfo{author}{\bibfnamefont{R.}~\bibnamefont{Roßbach}},
  \bibinfo{author}{\bibfnamefont{W.-M.} \bibnamefont{Schulz}},
  \bibinfo{author}{\bibfnamefont{M.}~\bibnamefont{Reischle}},
  \bibinfo{author}{\bibfnamefont{G.}~\bibnamefont{Beirne}},
  \bibinfo{author}{\bibfnamefont{M.}~\bibnamefont{Jetter}}, \bibnamefont{and}
  \bibinfo{author}{\bibfnamefont{P.}~\bibnamefont{Michler}},
  \bibinfo{journal}{J. Crystal Growth} \textbf{\bibinfo{volume}{in press}},
  \bibinfo{pages}{xxx} (\bibinfo{year}{2008}).

\bibitem[{\citenamefont{Brown and Twiss}(1956)}]{Hanbu}
\bibinfo{author}{\bibfnamefont{R.~H.} \bibnamefont{Brown}} \bibnamefont{and}
  \bibinfo{author}{\bibfnamefont{R.~Q.} \bibnamefont{Twiss}},
  \bibinfo{journal}{Nature} \textbf{\bibinfo{volume}{177}}, \bibinfo{pages}{27}
  (\bibinfo{year}{1956}).

\bibitem[{\citenamefont{Hermannst{\"a}dter
  et~al.}(2006)\citenamefont{Hermannst{\"a}dter, Beirne, Wang, Rastelli,
  Schmidt, and Michler}}]{Herma}
\bibinfo{author}{\bibfnamefont{C.}~\bibnamefont{Hermannst{\"a}dter}},
  \bibinfo{author}{\bibfnamefont{G.~J.} \bibnamefont{Beirne}},
  \bibinfo{author}{\bibfnamefont{L.}~\bibnamefont{Wang}},
  \bibinfo{author}{\bibfnamefont{A.}~\bibnamefont{Rastelli}},
  \bibinfo{author}{\bibfnamefont{O.~G.} \bibnamefont{Schmidt}},
  \bibnamefont{and} \bibinfo{author}{\bibfnamefont{P.}~\bibnamefont{Michler}},
  \bibinfo{journal}{AIP\ Conf.\ Proc.} \textbf{\bibinfo{volume}{893}},
  \bibinfo{pages}{875–6} (\bibinfo{year}{2006}).

\bibitem[{\citenamefont{Bayer et~al.}(2002)\citenamefont{Bayer, Ortner, Stern,
  Kuther, Gorbunov, Forchel, Hawrylak, Fafard, Hinzer, Reinecke
  et~al.}}]{Bayer2}
\bibinfo{author}{\bibfnamefont{M.}~\bibnamefont{Bayer}},
  \bibinfo{author}{\bibfnamefont{G.}~\bibnamefont{Ortner}},
  \bibinfo{author}{\bibfnamefont{O.}~\bibnamefont{Stern}},
  \bibinfo{author}{\bibfnamefont{A.}~\bibnamefont{Kuther}},
  \bibinfo{author}{\bibfnamefont{A.~A.} \bibnamefont{Gorbunov}},
  \bibinfo{author}{\bibfnamefont{A.}~\bibnamefont{Forchel}},
  \bibinfo{author}{\bibfnamefont{P.}~\bibnamefont{Hawrylak}},
  \bibinfo{author}{\bibfnamefont{S.}~\bibnamefont{Fafard}},
  \bibinfo{author}{\bibfnamefont{K.}~\bibnamefont{Hinzer}},
  \bibinfo{author}{\bibfnamefont{T.~L.} \bibnamefont{Reinecke}},
  \bibnamefont{et~al.}, \bibinfo{journal}{Phys.\ Rev.\ B}
  \textbf{\bibinfo{volume}{65}}, \bibinfo{pages}{195315}
  (\bibinfo{year}{2002}).

\bibitem[{\citenamefont{Vogel et~al.}(2007)\citenamefont{Vogel, Ulrich,
  Hafenbrak, Michler, Wang, Rastelli, and Schmidt}}]{Vogel}
\bibinfo{author}{\bibfnamefont{M.~M.} \bibnamefont{Vogel}},
  \bibinfo{author}{\bibfnamefont{S.~M.} \bibnamefont{Ulrich}},
  \bibinfo{author}{\bibfnamefont{R.}~\bibnamefont{Hafenbrak}},
  \bibinfo{author}{\bibfnamefont{P.}~\bibnamefont{Michler}},
  \bibinfo{author}{\bibfnamefont{L.}~\bibnamefont{Wang}},
  \bibinfo{author}{\bibfnamefont{A.}~\bibnamefont{Rastelli}}, \bibnamefont{and}
  \bibinfo{author}{\bibfnamefont{O.~G.} \bibnamefont{Schmidt}},
  \bibinfo{journal}{Appl.\ Phys.\ Lett.} \textbf{\bibinfo{volume}{91}},
  \bibinfo{pages}{051904} (\bibinfo{year}{2007}).

\end{thebibliography}
\end{document}